%% file: main.tex
\documentclass[twoside,leqno,twocolumn]{article}

\usepackage[letterpaper]{geometry}

\usepackage{ltexpprt}

\usepackage{bm}
\usepackage{bbm}
\usepackage{float}
\usepackage{subcaption}
\usepackage{xcolor}
\usepackage{graphicx}
\usepackage{amsfonts}
\usepackage{amsmath}
\usepackage{amssymb}
\DeclareMathAlphabet{\altmathcal}{OMS}{cmsy}{m}{n}
\usepackage[cal=cm]{mathalfa}
\usepackage{booktabs, arydshln}
\usepackage{multirow}
\usepackage{array}
\usepackage[font=small,skip=0.2cm]{caption}

\newcolumntype{H}{>{\setbox0=\hbox\bgroup}c<{\egroup}@{}}

\usepackage[ruled, algo2e, linesnumbered, noend]{algorithm2e}
\SetKw{KwBy}{by}

\newcommand{\model}{\textsc{$D^{2}M$}}
\newcommand{\method}{\textsc{AnomalyShield}}

\newcommand{\hide}[1]{}

\definecolor{as-color}{HTML}{FD8424}
\definecolor{rand-color}{HTML}{7FCD65}

\newtheorem{problem}{Problem}

\usepackage{enumitem}
\newlist{labeling}{enumerate}{1}

\newlist{itemdesc}{description}{1}
\setlist*[itemdesc,1]{font=\normalfont,leftmargin=15mm,labelwidth=*, labelindent=0.5cm}

\begin{document}
\setlength{\belowdisplayskip}{0pt} \setlength{\belowdisplayshortskip}{0pt}
\setlength{\abovedisplayskip}{0pt} \setlength{\abovedisplayshortskip}{0pt}

\title{\Large D$^{2}$M: Dynamic Defense and Modeling of Adversarial Movement in Networks} 
\author{Scott Freitas \thanks{Georgia Tech, Atlanta, GA (safreita@gatech.edu)}
\and Andrew Wicker \thanks{Research while at Microsoft. Now at Uber, Seattle, WA (andrew.wicker@uber.com)}
\and Duen Horng (Polo) Chau \thanks{Georgia Tech, Atlanta, GA (polo@gatech.edu)}
\and Joshua Neil \thanks{Microsoft Corp, Seattle, WA (joshua.neil@microsoft.com)}}

\date{}

\maketitle


\fancyfoot[R]{\scriptsize{Copyright \textcopyright\ 2020 by SIAM\\
Unauthorized reproduction of this article is prohibited}}





\begin{abstract} \small\baselineskip=9pt
Given a large enterprise network of devices and their authentication history (e.g., device logons), how can we quantify network vulnerability to lateral attack and identify at-risk devices? 
We systematically address these problems through \model{}, the first framework that models lateral attacks on enterprise networks using multiple attack strategies developed with researchers, engineers, and threat hunters in the Microsoft Defender Advanced Threat Protection group.
These strategies integrate real-world adversarial actions (e.g., privilege escalation) to generate attack paths: a series of compromised machines.
Leveraging these attack paths and a novel Monte-Carlo method, we formulate network vulnerability as a probabilistic function of the network topology,  distribution of access credentials and initial penetration point. 
To identify machines at risk to lateral attack, we propose a suite of five fast graph mining techniques, 
including a novel technique called \method{} inspired by node immunization research.
Using three real-world authentication graphs from Microsoft and Los Alamos National Laboratory (up to 223,399 authentications), we report the first experimental results on network vulnerability to lateral attack,
demonstrating \model{}'s unique potential to empower IT admins to develop robust user access credential policies.
\end{abstract}

\input{Sections/01-Introduction.tex}

\input{Sections/05-Related.tex}

\input{Sections/02-Overview.tex}

\input{Sections/03-Methodology.tex}

\input{Sections/04-Experiments.tex}

\input{Sections/07-Conclusion.tex}

\input{Sections/08-Acknowledgments}

\bibliographystyle{siamplain}
\bibliography{main.bib}

\end{document}

%% file: Sections/01-Introduction.tex
\section{Introduction}\label{section:intro}
Attack campaigns from criminal organizations and nation state actors are quickly becoming one of the most powerful forms of disruption. 
In 2016 alone, 
malicious cyber activity cost the U.S. economy between \$57 and \$109 billion \cite{council}. 
These cyber-attacks are often highly sophisticated, 
targeting governments and large-scale enterprises to interrupt critical services and steal intellectual property \cite{crowdstrike}.
Unfortunately, once an attacker has compromised a single credential for an enterprise machine, the
\textbf{whole network becomes vulnerable to lateral attack movements} \cite{hagberg2014connected}, allowing the adversary to eventually gain control of the network (i.e., escalating privileges via credential stealing \cite{duckwall2013hello}).

\begin{figure}[t!]
\centering
\includegraphics[width=0.85\linewidth]{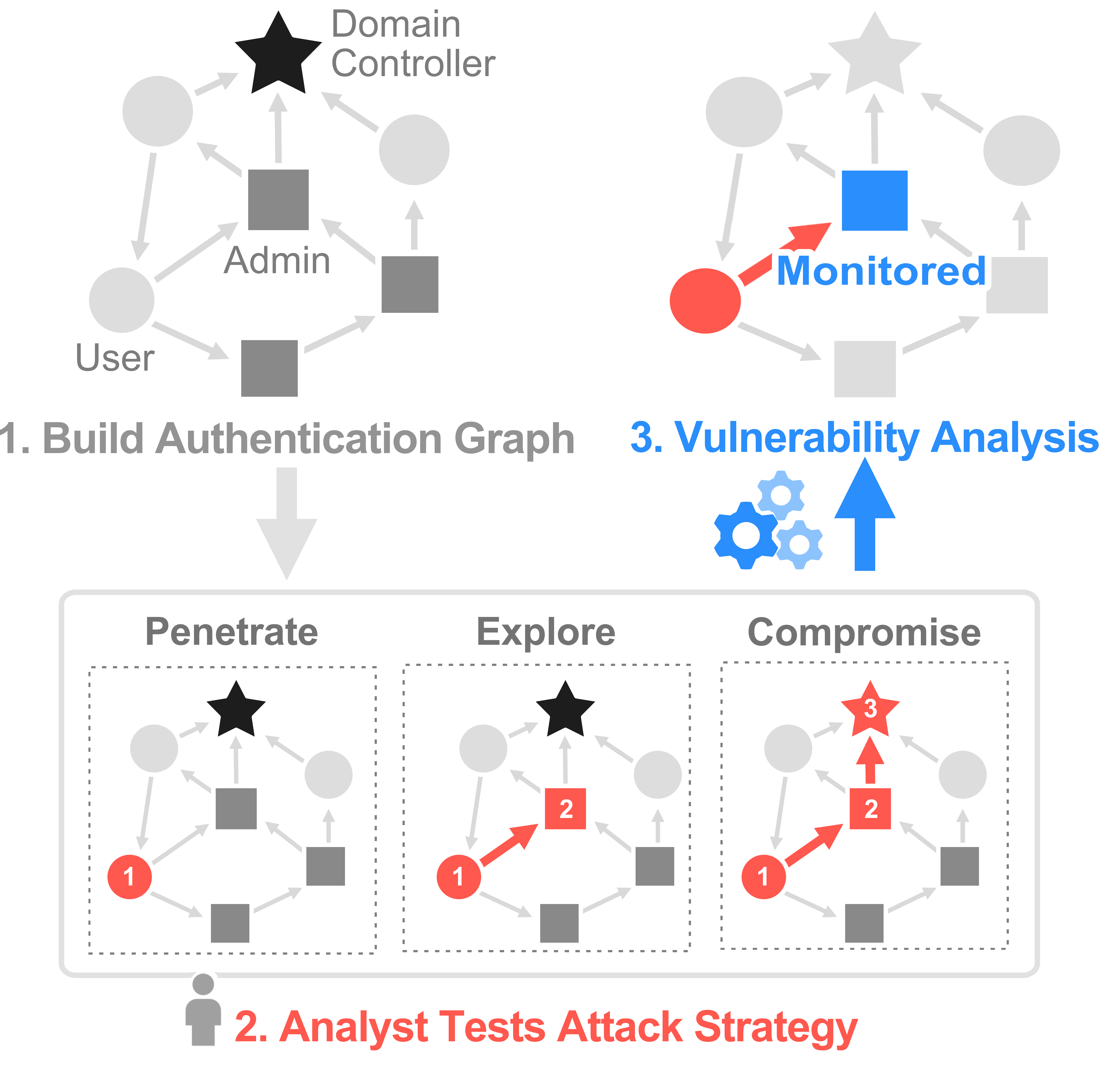}
\caption{Our \model{} framework: 
\textbf{1.} Builds an authentication graph from device authentication history;
\textbf{2.} Allows security analysts to test different attack strategies to study network vulnerability;
\textbf{3.} Identifies at-risk machines to monitor,
preempting lateral attacks.
}
\label{fig:crown}
\end{figure}

Despite their prevalence, observing and analyzing lateral attacks is challenging for multiple reasons:
(1) lateral attacks are still relatively sparse compared to the unsuccessful attack; 
(2) attack ground-truth is hard to ascertain, and generally partially uncovered through investigation; 
(3) incident reports are frequently withheld from the public for security and privacy concerns;
and (4) due to the fact that the adversary already has a valid credential for the network (e.g., gained through phishing \cite{banu2013comprehensive}), attackers can operate as a legitimate user. 
While real attack data does exist---due to the above challenges, 
it is rarely fully visible, or accessible,
making the study of a ``complete'' attack highly problematic.

\medskip
\noindent
\textbf{Our Contributions}

We propose \model{},
the first framework that systematically quantifies network vulnerability to lateral attack and identifies at-risk devices 
(Fig. \ref{fig:crown}).

Our major contributions include:

\begin{itemize}[topsep=1mm, itemsep=0mm, parsep=1mm, leftmargin=*]
    
    \item \textbf{Attack Strategies}
    \model{} enables security researchers to integrate their crucial domain knowledge from studying prior attacks in the form of attack strategies.
    We developed
    three attack strategies by actively engaging researchers, engineers and threat hunters in the Microsoft Advanced Threat Protection group,
    whose expertise lies in tracking down adversaries in a post-breach environment (once adversary is on  network). 
    \model{} integrates real-world adversarial actions (e.g., privilege escalation), generating attack paths consisting of a series of compromised machines (Sec. \ref{section:simulation}; Fig. \ref{fig:crown}.2).

    \item \textbf{Network Vulnerability Analysis} 
    We formulate a novel Monte-Carlo method for lateral attack vulnerability 
    as a probabilistic function of the network topology, distribution of access credentials and initial penetration point (Fig. \ref{fig:crown}.3). 
    This empowers IT admins to develop robust user access credential policies and
    enables security researchers to study the vulnerability of a network to lateral attack 
    (Sec.~\ref{section:vulnerability}).
    
    \item \textbf{Network Defense by Identifying At-risk Machines} 
    To identify machines at risk to lateral attack, we propose a suite of five fast graph mining techniques, 
    including a novel technique called \method{} which prioritizes machines with anomalous neighbors and high eigencentrality (Fig. \ref{fig:crown}.3; Sec.~\ref{sec:defense}). 
    
    \item \textbf{Evaluation Using Real-World Data}
    Using three real-world authentication graphs from Microsoft and Los Alamos National Laboratory (LANL; up to 223,399 authentications), we report the first experimental results on network vulnerability to lateral attack and at-risk machine identification (Sec.~\ref{section:simulation}).

    \item \textbf{Impact to Microsoft and Beyond.} 
    The Microsoft Defender Advanced Threat Protection product is deployed to thousands of enterprises around the world, and is a leader in the Endpoint Detection and Response (EDR) market~\cite{lefferts2019gartner}. 
    The ability to detect and prevent lateral movement is one of the most challenging areas of post-breach detection. 
    This research has led to major impact to Microsoft products, inspiring changes to the product's approach to lateral movement detection.
    
\end{itemize}

Table~\ref{table:notation} describes the main symbols used in the paper. 
We follow standard notation and use capital bold letters for matrices (e.g., $\bm{A}$), lower-case bold letters for vectors (e.g., $\bm{a}$) and calligraphic font for sets (e.g., $\altmathcal{S}$). 

%% file: Sections/05-Related.tex
\section{Background and Our Differences}
Our work intersects the domains of lateral attack and graph mining, we briefly review related work below.
Different from existing work that detects lateral movement after an adversary is on the network, 
our work \textbf{quantifies network vulnerability to lateral attack} and \textbf{identifies at-risk machines}. 
Another important distinction is that this work uses real-world enterprise authentication graphs, 
while most prior work has not.

\input{Tables/Notation.tex}

\subsection{Detecting Lateral Attacks}
Significant research in \textit{detecting} lateral movement in networks has been done \cite{liu2018latte, neil2013scan, noureddine2016game, fawaz2016lateral}. 
Latte \cite{liu2018latte}, a graph based detection framework, discovers potential lateral movement in a network using forensic analysis of known infected computers.
In \cite{neil2013scan}, Neil et al. detects lateral attacks using statistical detection of anomalous graph patterns  (e.g., paths, stars) over time. 
Alternatively, Noureddine et al. \cite{noureddine2016game} proposes a zero-sum game to identify which machines a defender should monitor to slow down an attacker. 
Finally, a data fusion technique is proposed by Fawaz et al. \cite{fawaz2016lateral}, where host-level process communication graphs are aggregated into system-wide communication graphs to detect lateral movement.

\subsection{Graph Mining \& Network Security}


Graph mining has been extensively applied to the more general domain of network security.
Authentication graphs have been used to study network security from a variety of viewpoints \cite{hagberg2014connected, kent2015authentication, neil2013scan}. 
In \cite{hagberg2014connected}, Hagberg et al. studies credential hopping in authentication graphs and finds that by reducing a machine's credential cache, lateral movement can be restricted. 
Alternatively, Kent et al. \cite{kent2015authentication} develops individual user authentication graphs to differentiate normal authentication activity from malicious. 
Orthogonal to the authentication graph and \textit{our work}, attack graphs have been proposed to analyze a network's risk to known security issues \cite{sheyner2003tools, ammann2002scalable, jha2002two}. These graphs represent sequences of known system vulnerabilities that can be maliciously exploited; and are often used by IT admins to determine patch priority.

%% file: Tables/Notation.tex
\begin{table}[t]
\small
\centering
 \begin{tabular}{l l} 
 \toprule
 \textbf{Symbol} & \textbf{Definition} \\
 \midrule
 $G$ & Directed, unweighted, attributed graph \\
 $\altmathcal{V}$, $\altmathcal{E}$ & Set of nodes and edges in graph $G$ \\
 $n, m$ & Number nodes $|V|$, edges in $|\altmathcal{E}|$ in $G$ \\ 
 $\bm{A}(i, j)$ & Adj. matrix of $G$ at \textit{i}th row, \textit{j}th column \\
 $\bm{u}(i)$ & Eigenvector at position $i$ \\
 $\altmathcal{C}, c$ & Credential set; credential instance \\
 $\textit{D}$ & Credential generation process \\
 $\bm{d}$ &  Credential vector \\
 $\altmathcal{H}$, $h$ & Ordered hygiene set; hygiene instance \\
 $N^{+}(v)$, $N(v)$ & Successors of $v$; neighbors of $v$ \\
 $\altmathcal{R}$, $\altmathcal{T}$ & Set of start nodes; set of attacker moves \\
 $\altmathcal{S}_k$ & Set of $k$ nodes to monitor \\
 $SV(\altmathcal{S}_k)$, & Shield value of $\altmathcal{S}_k$ \\
 $AV(\altmathcal{S}_k)$ & Anomaly value of $\altmathcal{S}_k$ \\
 $L(G)$ & Vulnerability of $G$ to lateral attacks \\
 $\bm{p}$ & Attack path \\
 $\bm{a}$ & Per-machine anomaly vector \\
 $i_s$ & Number of sub-path intervals \\
 $k$ & Number of machines to vaccinate \\

\bottomrule
\end{tabular}
\caption{Symbols and Definition
}
\label{table:notation}
\end{table}

%% file: Sections/02-Overview.tex
\section{Authentication Graph}\label{section:overview}
\model{} converts authentication history of network devices into an \textit{authentication graph}, where directed edges represent machine-machine authentications (i.e., logons) in an organization. 
Below, we provide an overview of the authentication graph setup and the infusion of real-world domain knowledge into its construction.

\subsection{Building Graph Structure}\label{authentication_graph}
Modern enterprise computer networks typically rely on one of two types of centrally managed authentication mechanisms to authenticate user activity: Microsoft NTLM \cite{ntlm} or MIT Kerberos \cite{neuman1994kerberos}. To avoid repeated authentication with network resources (e.g., printer, corporate web sites, email), both NTLM and Kerberos implement credential caching where user credentials are stored on the computer until either the user logs off (Kerberos), or the machine is restarted (NTLM) \cite{hagberg2014connected}. While these cached credentials are convenient for legitimate user activity, they pose significant risk for malicious exploitation \cite{duckwall2013hello, soria2017detecting}.

Leveraging this authentication history, we form a directed, unweighted graph $G=(\altmathcal{V},\altmathcal{E})$, where an edge represents an authentication between source machine $v_s$ and destination machine $v_d$ (see Fig.~\ref{fig:crown}.1).
We combine all authentications between two machines into a single edge. 
These authentication events are recorded over a period of time, forming the graph topology of an organization \cite{hagberg2014connected, kent2015authentication}. 
To verify that a remote connection between two machines can be established, authentication information is passed using cached credentials. 
In an enterprise network, these credentials typically follow a hierarchical scheme: \textit{user} ($c_1$) at the bottom, \textit{local admin} ($c_2$) and \textit{network admin} in the middle ($c_3$), and \textit{domain admin} ($c_4$) at the top ($c_1 < c_2 < c_3 < c_4$) \cite{soria2017detecting}.
Depending on the type of cached credential, it will be valid until the user logs out (Kerberos) or until the machine is restarted (NTLM). 

\subsection{Integrating Domain Knowledge}
To enhance \model{} with realistic security and attack practices, we integrate the following three components into our framework:
(1) per-machine \textbf{credential caching}; 
(2) \textbf{network hygiene} (i.e., how many `users' and `admins' on the network); and 
(3) \textbf{domain controller} modeling. 

\vspace{2mm}
\noindent
\textbf{Credential Caching}
We embed attribute information into graph $G$ by giving each machine $v\in V$ a cached credential. These credentials are stored as a vector $\bm{d}\in \mathbb{R}^n$, where each entry is a machine in the authentication graph containing the most recent credential  $\bm{d}(i)=c$. 
While some credential schemes have additional levels and queue lengths as active directory policies, 
our approach captures representative security information. 

\vspace{2mm}
\noindent
\textbf{Network Hygiene} 
We model various credential distributions through three levels of hygiene $h \in \altmathcal{H}$ due to the unavailability of credential information in the network $\bm{d} = <c_1,c_2,...,c_n>$ where $n = |\altmathcal{V}|$.  Each hygiene level ($h_1$: low, $h_2$: medium, $h_3$: high) represents the frequency with which credential types are observed on the network. Intuitively, a low hygiene level ($h_1$) models a network with loose IT policies and an abundance of high-level administrator credentials. In contrast, a high hygiene level ($h_3$) represents a network with strict IT policies and limited distribution of admin credentials. We select each hygiene distribution $h\in \altmathcal{H}$ as: $h_1=\{c_1$: $n$, $c_2$: $n/2$, $c_3$: $n/5$, $c_4$: $n/20\}$, $h_2=\{c_1$: $n$, $c_2$: $n/4$, $c_3$: $n/10$, $c_4$: $n/50\}$ and $h_3=\{c_1$: $n$, $c_2$: $n/8$, $c_3$: $n/20$, $c_4$: $n/80\}$, which are determined experimentally in conjunction with domain experts.

In practice, we distribute these credentials for a given hygiene $h$ as follows. 
For every machine in the network $v\in \altmathcal{V}$ we assign the lowest authorization level $\bm{d}(v)$ = $c_1$. 
We then distribute higher level credentials as follows---for each increasing credential level $c\in \{c_2, c_3, c_4\}$, we randomly select $h(c)$ machines from $\altmathcal{V}$ and loop through each one, replacing it's credential level with a higher one. 
While these distributions cannot match every organization's IT policies, we select them to model a broad range.

\vspace{2mm}
\noindent
\textbf{Domain Controller \& Privilege Escalation} The final component we model is the domain controller, which controls access to network resources. When a source machine $v_s$ attempts to establish a remote connection to a destination machine $v_d$, the domain controller determines if $v_s$ has sufficient privileges $\bm{d}(v_s) \geq \bm{d}(v_d)$.
Since an organization's domain controller(s) are never observed with certainty, we identify it using PageRank ($\alpha$=0.15) \cite{page1999pagerank}---assigning the machine with largest PageRank vector $\bm{r}\in \mathbb{R}^n$ component the role of domain controller $v_{dc}$ = $argmax(\bm{r})$. After discussions with domain experts, we make the simplifying assumption that the machine with largest PageRank is the domain controller $v_{dc}$, since it often has the largest number of incoming edges (from incoming authentication requests).

Finally, we incorporate the concept of privilege escalation by allowing the attacker to connect to a machine that is one credential level higher. That is, if the attacker has collected credentials $c_1$ and $c_2$, they can connect to a $c_1$, $c_2$, or $c_3$ machine. 
In practice, this is done through mining the memory of the machine to gain higher levels of credential \cite{mulder2016mimikatz}.

\section{Formulating the Research Problems}\label{sec:formulation}
We formally define the three problems that \model{} addresses below.
Then we present our solutions for them in 
\textbf{Section 
\ref{section:simulation}},
\textbf{\ref{section:vulnerability}},
and
\textbf{\ref{sec:defense}}
respectively.

\begin{problem}\label{problem:1}\textbf{Lateral Attack Modeling}

\begin{description}[topsep=-1mm, itemsep=0mm, parsep=1mm, leftmargin=6mm, itemindent=0mm]

\item [Given:] an attack strategy, an initial penetration point, and directed unweighted graph $G$ with associated credential distribution $\bm{d}\in D$

\item [Find:] an attack path $\bm{p}=<v_1, v_2,...v_i...,v_t>$ in graph $G$ that starts from the penetration point and reaches the domain controller, while escalating privileges in increasing order (see Fig.~\ref{fig:attack_path})

\end{description}

\end{problem}

\begin{problem}\label{problem:2}\textbf{Lateral Attack Vulnerability}

\begin{description}[topsep=-1mm, itemsep=0mm, parsep=1mm, leftmargin=6mm, itemindent=0mm]

\item [Given:] graph $G$ with credential distribution $\bm{d}\in D$

\item [Measure:] vulnerability $L(G)$ to lateral attacks

\end{description}

\end{problem}

\begin{problem}\label{problem:3}\textbf{Lateral Attack Defense}

\begin{description}[topsep=-1mm, itemsep=0mm, parsep=1mm, leftmargin=6mm, itemindent=0mm]

\item [Given:]  graph $G$ with credential distribution $\bm{d}\in D$, and suspected adversary movement $\bm{p}$ 

\item [Identify:]  $k$ best machines to monitor for  attacks

\end{description}

\end{problem}

%% file: Sections/03-Methodology.tex
\section{
D\textsuperscript{2}M: 
Lateral Attack Modeling
}
\label{section:simulation}

We present our solution for the \textit{lateral attack modeling} problem (Sect.~\ref{sec:formulation}: Problem~\ref{problem:1}). 
We begin with an overview of the lateral attack process in Section~\ref{sec:modeling-attacks}. 
Section~\ref{attack_models} presents lateral attack strategies---developed with Microsoft domain experts---that produce lateral movement. 
Section~\ref{section:attack_sim} details the algorithm for modeling lateral attacks on authentication graphs.

\subsection{Lateral Attack Overview}\label{sec:modeling-attacks}
An enterprise attack typically follows a kill chain, which can be distilled into three phases---(1) penetration of the network; (2) exploration of the network and escalation of privileges; and (3) exfiltration of data back to the command and control server \cite{sexton2015attack}. We discuss each phase below and highlight our modeling assumptions.

\smallskip
\noindent
\textbf{Penetration} An enterprise network is typically penetrated through two mechanisms---(a) phishing campaigns targeting organization employees or (b) incidental exposure from employees downloading malware on high-risk websites (drive-by download) \cite{le2013anatomy}. We assume the former, since sophisticated adversaries often target enterprise networks for penetration. A phishing campaign begins by targeting organization employees through authentic looking emails containing malicious attachments or web links. These malicious attachments contain malware that installs a backdoor; once a backdoor is installed the attacker gains remote access to the machine, penetrating the enterprise network.
We model this penetration process by assuming that most compromised employees (machines) $v\in \altmathcal{V}$ are at the $c_1$ credential level and let the attacker randomly start on any of these machines $\altmathcal{R}=\{v\in \altmathcal{V} \mid \bm{d}(v) = c_1\}$.

\begin{figure}[t]
\centering
\includegraphics[width=0.85\linewidth]{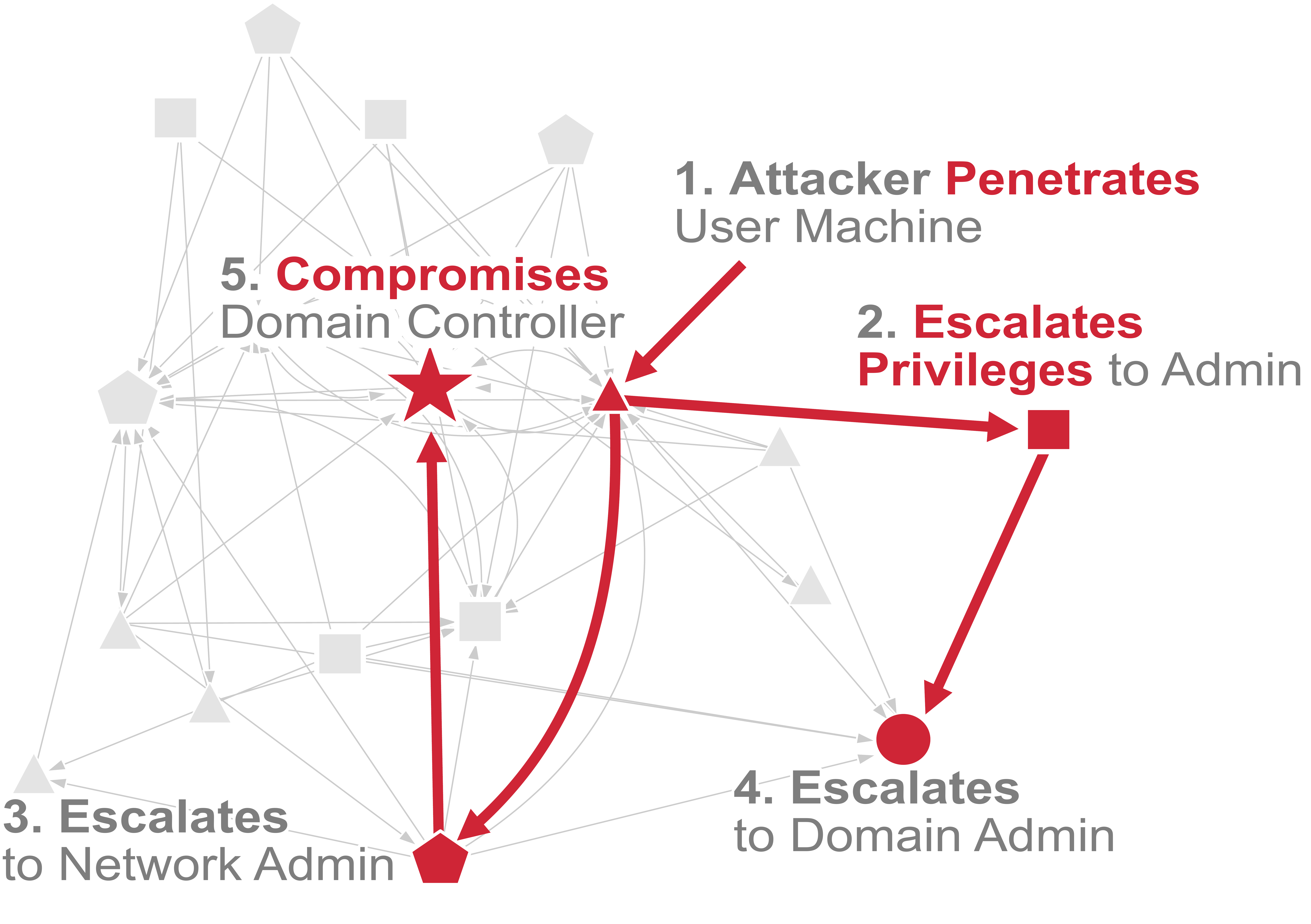}

\caption{Attack path generated by \model{}. 
\textbf{1.} Network is penetrated; 
\textbf{2}-\textbf{4.} Attacker explores the network and escalates privileges; 
\textbf{5.} Attacker compromises the domain controller, gaining control of the network
}
\label{fig:attack_path}
\end{figure}

\smallskip
\noindent
\textbf{Explore \& Exploit} Once an adversary is on a network, their goal is to explore the network and escalate privileges. This process begins by stealing the infected machines cached credentials, allowing them to authenticate with neighboring machines. These credentials can be stolen in a number of ways, however, it is beyond the scope of this work and we refer the reader to \cite{soria2017detecting}. Once the adversary has connected to a neighboring machine, they again steal the cached credentials \cite{duckwall2013hello} and continue this process until they have obtained domain admin privileges $c_4$. We model this attack process in two ways---(1) black-box, where the attacker has no prior information on the network (i.e., normal pattern of authentications); and (2) gray-box, where the attacker has prior information on the network layout, possibly through prior reconnaissance or inside help. 

\smallskip
\noindent
\textbf{Exfiltrate} After the adversary has obtained a domain admin credential $c_4$, they're able to connect to any networked machine, freely exploring the network until they reach the domain controller. Upon accessing the domain controller, the attacker gains full control over the network. At this point the adversary can sweep the network for valuable information and exfiltrate with impunity. We leave modeling this aspect of the kill chain to future work.

\subsection{Lateral Attack Strategies}\label{attack_models}
In conjunction with domain experts, we develop three attack strategies to model lateral attacks on authentication graphs; one black-box and two gray-box. 

\subsubsection{Black-Box Attack}
In the black-box setting we assume the attacker has no knowledge about the network and model movement through a modified random walk called \textit{RandomWalk-Explore} (RWE). 

\smallskip
\noindent
\textbf{RandomWalk-Explore (RWE)} with 0.85 probability draws a machine $v$ uniformly at random from the set of unvisited neighboring machines $\altmathcal{T}$. With probability 0.15, the attacker randomly jumps with uniform probability to a machine in $\altmathcal{R}$; this helps to model some of the usual behavior that can occur during an attack (e.g., when an attacker finds remote machine information in plain-text). In addition, we select 0.15 as the random jump probability to align with information retrieval literature \cite{page1999pagerank}. We model the RWE process in Equation~\ref{eq:move_RWE}, which describes the probability mass function (PMF) of a discrete random variable $X_{1}$, which can take on any value in the range $R_{X_{1}}=\altmathcal{T} \cup \altmathcal{R}$ with probability $P_{X_1}(v)$.

\begin{equation}\label{eq:move_RWE}
\small
    P_{X_1}(v) = 
    \begin{cases}
        0.15 / |\altmathcal{R}|,     & \text{if } v\in \altmathcal{R} \\
        0.85 /|\altmathcal{T}|,   & \text{if } v\in \altmathcal{T} \\
        0,              & \text{otherwise}
    \end{cases}
\end{equation}

\subsubsection{Gray-Box Attacks}
In the gray-box setting, the attacker has additional information in the form of the network topology---allowing for informed attack strategies. We propose two strateiges, \textit{Rank-Explore} (RE) and \textit{Degree-Explore} (DE). 

\smallskip
\noindent
\textbf{Rank-Explore (RE)} with 0.85 probability draws a machine $v$ at random from the set of unvisited neighboring machines $\altmathcal{T}$ with weight proportional to its PageRank vector $r$. With probability 0.15, the attacker randomly jumps with uniform probability to a machine in $\altmathcal{R}$. This process is modeled in Equation~\ref{eq:move_RE}.

\begin{equation}\label{eq:move_RE}
\small
    P_{X_2}(v) = 
    \begin{cases}
        0.15/|\altmathcal{R}|,           & \text{if } v\in \altmathcal{R} \\
        0.85 \cdot \bm{r}(v) / \sum\limits_{i\in \altmathcal{T}}\bm{r}(i), & \text{if } v\in \altmathcal{T} \\

        0,              & \text{otherwise}
    \end{cases}
\end{equation}

\smallskip
\noindent
\textbf{Degree-Explore (DE)} with 0.85 probability draws a machine $v\in \altmathcal{T}$ with weight proportional to the distribution of the network's degree vector $\bm{\delta} = diag(\bm{A}\cdot \bm{e})$. With probability 0.15, the attacker randomly jumps with uniform probability to a machine in $\altmathcal{R}$. This process is modeled in Equation~\ref{eq:move_DE}.

\begin{equation}\label{eq:move_DE}
\small
    P_{X_3}(v) = 
    \begin{cases}
        0.15/|\altmathcal{R}|,           & \text{if } v\in \altmathcal{R} \\
        0.85 \cdot \bm{\delta}(v) / \sum\limits_{i\in \altmathcal{T}}\bm{\delta}(i), & \text{if } v\in \altmathcal{T} \\

        0,              & \text{otherwise}
    \end{cases}
\end{equation}

After a neighbor $v$ has been selected by the attack strategy, we check that the attacker has the required credential level to visit this machine. For example, if $c_2$ is the current highest collected credential, then the attacker can move to any machine with credential level $c_1$, $c_2$, or $c_3$.
If the move is valid, we update the set of unvisited neighbors $\altmathcal{T}$ according to Equation~\ref{eq:move_update} and allow the attacker to collect that machine's credential.

\begin{equation}\label{eq:move_update}
\small
    \altmathcal{T} = \altmathcal{T} \setminus \{v\} \cup N^{+}(v)
\end{equation}

\subsection{Lateral Attack Algorithm}
\label{section:attack_sim}
We allow the attacker to randomly penetrate various points of the network ($v\in R$) 
and then move according to one of the three strategies: RWE, RE and DE, until the domain controller $v_{dc}$ is reached or there are no neighbors to visit. 
Each successful run of this simulation generates an attack path $\bm{p}=<v_1, v_2,...v_i...,v_{dc}>$, representing the sequence of machines visited, with the last node $v_{dc}$ representing the domain controller. This process is modeled in Algorithm~\ref{algorithm:attack} and repeated for multiple credential distributions $\bm{d}\in D$ to eliminate bias from a single distribution. An example attack path generated from Algorithm~\ref{algorithm:attack} can be seen in Figure~\ref{fig:attack_path}. 

\input{Algorithms/Adversary.tex}

\subsection{Analysis of Lateral Attack Algorithm}
The time and space complexity of Algorithm~\ref{algorithm:attack} is $O(n^2)$ and $O(n + m)$, respectively.

There are two time expensive computations, PageRank $O(n)$; and attack strategy machine selection inside the while loop $O(n)$. Since the while loop can visit every node in the graph, the worst case complexity will be $O(n^2)$. Space is linear with respect to nodes and edges $O(n + m)$ in the graph.
Detailed proofs are omitted to save space.

\section{
D\textsuperscript{2}M: Lateral Attack Vulnerability}\label{section:vulnerability}

We present our solution for the \textit{lateral attack vulnerability} problem (Sect.~\ref{sec:formulation}: Problem~\ref{problem:2}). 
We begin by discussing the importance of network vulnerability scoring. 
We then formally introduce our method of measuring a network's vulnerability to lateral attacks. Finally, we discuss alternative graph vulnerability scores and why they are less suited to the task of measuring vulnerability to lateral movement.

\smallskip
\noindent
\textbf{Vulnerability Scoring}
To make data driven decisions regarding IT policy in an enterprise network, it is important to quantify the risk a network faces to lateral movement. Unfortunately, directly measuring this risk is difficult, requiring complex interactions of many unknown variables. To simplify these interactions, we propose to quantify network vulnerability to lateral attack $L(\cdot)$ as a function of three random variables---(1) network topology $G$,
(2) distribution of credentials $\bm{d} \in D$ and (3) initial point of penetration $v\in \altmathcal{R}$. 

Since the true credential distribution $\bm{d} = <c_1, c_2, c_3, c_4>$ is unknown, along with knowledge of the organizations IT policies (strict, loose: Section~\ref{authentication_graph}), we model credential distributions through the use of hygiene levels $h\in \altmathcal{H}$. 
For a given hygiene level $h\in \altmathcal{H}$, we can marginalize out the dependency of the vulnerability score to the credential distribution $\bm{d}\in D_h$ in expectation, reducing the vulnerability score to $L(G, \altmathcal{H}=h, V=v)$. In addition, we can simulate the attacker penetrating many different points in the network $v\in \altmathcal{R}$,
allowing us to marginalize out the dependency to $v$ and reduce the score to $L(G, h)$. We can view this process in Equation~\ref{eq:vulnerability_1} through the lens of Monte Carlo simulation, where in expectation we compute the graph vulnerability across many different credential distributions $\bm{d}\in D$ and start nodes $v\in \altmathcal{R}$.

\begin{equation}\label{eq:vulnerability_1}
\small
    L(G, h) =  \frac{1}{|D_h}\frac{1}{|\altmathcal{R}|} \sum_{\bm{d}\in D_h}
     \sum_{v\in \altmathcal{R}}
     f(G, \bm{d}, v)
\end{equation}

The vulnerability score $L(G, h)$ is a real number between $0 \leq L(G, h) \leq 1$, where a higher value indicates a more vulnerable network for the given topology $G$ and hygiene level $h$. 
Intuitively, this score is saying that a network is more vulnerable if attacks are on average more successful for many credential distributions $\bm{d} \in D$ and penetration points $v\in \altmathcal{R}$.
We measure an attack's success through $f(\cdot)$, which simulates an attack using Algorithm~\ref{algorithm:attack}. A value of $f(G, \bm{d}, v) = 1$ indicates a successful attack, which we define as being able to reach the domain controller $v_{dc}$. Future work could generalize this to other targets such as high value servers.

We further simplify the vulnerability score $L(\cdot)$ by marginalizing out the dependency to hygiene level $h\in \altmathcal{H}$. This simplifies Equation~\ref{eq:vulnerability_1} to a function of the network topology $G$, as seen in Equation~\ref{eq:vulnerability_2}.

\begin{equation}
\small
\label{eq:vulnerability_2}
    L(G) = \sum_{h_i\in \altmathcal{H}} p(h_i) \cdot L(G, h_i)
\end{equation}

With no prior knowledge on the true distribution of hygiene levels in an organization, we assume a uniform prior $p(h) = 1/3$.

\smallskip
\noindent
\textbf{Alternative Scoring}
Significant work has gone into measuring the vulnerability of graphs \cite{chen2015node, saha2015approximation, tong2010vulnerability}. For example, in \cite{tong2010vulnerability} the authors define vulnerability of an undirected graph $G$ as the largest eigenvalue $L(G)\triangleq\lambda$ of the adjacency matrix. The intuition is that as the largest eigenvalue increases, so does the path capacity of the graph. 
However,
this form of topological vulnerability scoring can only indirectly measure the vulnerability of the graph to lateral movement since no security domain knowledge is integrated. 

\section{
D\textsuperscript{2}M: 
Lateral Attack Defense}
\label{sec:defense}

We present our solution for the \textit{lateral attack defense} problem (Sec.~\ref{sec:formulation}: Problem~\ref{problem:3}), where the objective is to identify the best set of $k$ machines  $\altmathcal{S}_k$ to monitor for lateral attacks. 
Once this set of machines $\altmathcal{S}_k$ has been identified, multiple safeguards can be implemented, including: changing the sensitivity of on device machine learning models and force resetting the password. 

We make the following assumptions during the defense process---(a) there exists per-machine anomaly detection models that alert on unusual behavior (e.g., deviation in port or process activity). Since behavioral deviations have a larger false positive rate, their behavior is anomalous but not necessarily malicious. For this reason, anomaly alerts are ill-suited for investigation in isolation due to low confidence. However, these deviation scores are useful for machine monitoring decisions, especially when these alerts aggregate together \cite{neil2013scan}. (b) We assume that each anomaly detection model is providing real-time feedback to the defender; and (c) that the defender views all anomalous activity as it occurs through the system alerts. While assumption (c) is strong, we leave it to future work to model partial information defense strategies.

\subsection{Defense Strategies}
\label{vaccination_models}

We propose a suite of five defense strategies, three \textit{static} and two \textit{dynamic}.
A \textit{static} strategy takes into account only the network topology $G$; useful for protecting machines when monitoring resources are limited.
A \textit{dynamic} strategy considers both the network topology $G$ and suspected lateral path movement activity $\bm{p}^{t}, \bm{p}^{t-1},...\bm{p}^{i}...,\bm{p}^{0}$, 
where $p^{i} \in \mathbb{R}^n$ is a sub-path containing suspicious activity in a given interval. 
This could be useful for real-time protection malicious activity investigation.

Each attack path $\bm{p}$ is divided into $i_s$ sub-paths, where each sub-path $\bm{p}^i$ is of equal size (except for, possibly, the last sub-path $p^t$) where $t\in [0, \lceil\frac{\bm{p}}{i_s}\rceil]$. 
A larger value of $i_s$ creates a few long sub-paths, which could represent fast moving attacks in the network; conversely, a small $i_s$ creates many short sub-paths, representing slow attacks.

\smallskip
\noindent
\textbf{Rank-Defense (RD)} statically identifies at-risk machines based on the network's PageRank \cite{page1999pagerank}. Assuming a sorted PageRank vector, we identify machines as follows: $\altmathcal{S}_k = \cup_{i=1}^k \bm{r}_i$.

\smallskip
\noindent
\textbf{Degree-Defense (DD)} statically vaccinates the network according to the machines in the network with highest degree. With a sorted degree vector, we identify machines as follows: $\altmathcal{S}_k = \cup_{i=1}^k \bm{\delta}_i$.
While RD and DD are simple defensive strategies, we are not aware of any work proposing to identify at-risk machines to lateral attacks using them.

\smallskip
\noindent
\textbf{NetShield (NS)} \cite{tong2010vulnerability} statically vaccinates the network according to the machine's Shield-Value ($SV$) in Equation~\ref{eq:NetShield_defense}. The actual selection of $\altmathcal{S}_k$ occurs in conjunction with the NetShield algorithm from \cite{tong2010vulnerability}, where the intuition is to select nodes with highest eigencentrality \cite{newman2016mathematics} 
while enforcing distance between selected machines (small or zero $\bm{A}(i, j$)). Here, $\bm{A}\in \{0, 1\}^{n\times n}$, $\lambda$ is the largest eigenvalue, and $\bm{u}$ is the associated eigenvector.

\begin{equation}\label{eq:NetShield_defense}
\small
    SV(\altmathcal{S}_k) = \sum_{i\in \altmathcal{S}_k} 2\lambda \cdot \bm{u}(i)^2 - \sum_{i, j\in \altmathcal{S}_k}\bm{A}(i,j) \bm{u}(i) \bm{u}(j)
\end{equation}

\smallskip
\noindent
\textbf{Random Anomalous Neighbor Defense (RAND)}
dynamically identifies machines by selecting an anomalous machine $v_a$ with weight proportional to its anomaly score $\bm{a}(v_a)$, where each element $\bm{a}(v) \in [0, 1]$ and $\bm{a}\in \mathbb{R}^n$. We assume that when an alert is generated for a machine in a sub-path, it produces a value of $\bm{a}(v) = 1$, repeating for every machine $v\in \bm{p}^i$. After machine monitoring set $\altmathcal{S}_k$ is identified using sub-paths $\bm{p}^i,...\bm{p}^0$, the anomaly scores are decayed $\bm{a}^{t+1} = \bm{a}^t / 2$ to give weight to recent activity (determined experimentally). 

The RAND strategy in described through Equations~\ref{eq:ran_defense} and \ref{eq:ran_defense2}. 
Eq.~\ref{eq:ran_defense} describes the PMF of discrete random variable $X_{4}$, which can take on any value in the range $R_{X_4}=\{v\in \altmathcal{V} \mid \bm{a}(v) > 0 \}$ with probability $P_{X_4}(v)$. After drawing a machine $v_a \sim X_4$, we uniformly at random select a neighbor from $v_a$. This can be seen in Equation~\ref{eq:ran_defense2}, which describes the PMF of discrete random variable $X_{5}$, where $X_5$ can take on any value in the range $R_{X_5} = N^+(v_a)$ with probability $P_{X_5}(v)$. This process repeats until $k$ machines have been selected.

\begin{equation}\label{eq:ran_defense}
\small
\begin{aligned}
    P_{X_4}(v) = 
    \begin{cases}
        \bm{a}(v) / \sum\limits_{i\in \altmathcal{V}}\bm{a}(i), & \text{if } v\in R_{X_4} \\

        0,              & \text{otherwise}
    \end{cases}    
\end{aligned}
\end{equation}

\begin{equation}\label{eq:ran_defense2}
\small
\begin{aligned}
    P_{X_5}(v) = 
    \begin{cases}
        1 / |N^+(v_a)|, & \text{if } v\in N^+(v_a) \\
        0,              & \text{otherwise}
    \end{cases}    
\end{aligned}
\end{equation}

\smallskip
\noindent
\textbf{\method{} (AS)}, a novel method we introduce for dynamic machine identification. 
We select machines for monitoring according to their \textsc{AnomalyValue} ($AV$) in Equation~\ref{eq:as_defense}, in combination with \method{} (Algorithm~\ref{algorithm:anomaly_shield}). The intuition is that we prioritize machines with anomalous neighbors and high eigencentrality. 

\begin{equation}\label{eq:as_defense}
\small
\begin{aligned}
    AV(\altmathcal{S}_k) &= \sum_{i\in \altmathcal{S}_k} \bm{u}(i) \sum_{j\in N(i)} \bm{a}(j)\bm{u}(j) \\
\end{aligned}
\end{equation}

Since both NetShield and AnomalyShield use eigenvector centrality as the underlying centrality metric, we convert the directed authentication graphs to undirected ones for use in the strategies.

\input{Algorithms/AnomalyShield.tex}

\subsection{Analysis of Defense Strategies}
We evaluate time and space complexity with respect to each strategy since they are the dominating defense cost. The space is uniform across strategy $O(n + m + k)$, with time complexity shown below. Proofs omitted for space.

\begin{equation}\label{eq:defense-time}
Time=
\small
\begin{cases}
    O(nlogn),           & \text{if defense = RD} \\
    O(nlogn),           & \text{if defense = DD} \\
    O(nk^2 + m),           & \text{if defense = NS \cite{chen2015node}}  \\
    O(kn + m),           & \text{if defense = AS} \\
    O(kn),           & \text{if defense = RAND} \\
\end{cases}
\end{equation}

%% file: Algorithms/Adversary.tex
\begin{algorithm2e}[t]
\footnotesize
    \DontPrintSemicolon
    \KwIn{Adj. matrix $\bm{A}$, $h$, attack strategy}
    \KwResult{Attack pattern $\bm{p}$}

    let $\bm{r}_o$ = PageRank($\bm{A}$);
    and $\bm{\delta}_o$ = diag($\bm{A} \cdot \bm{1}$)\;
    let $\bm{d} \sim D_h$ \tcp*{distribute credentials} 
    $\altmathcal{R} = \{v\in V \mid \bm{d}(v) = c_1\}$ \tcp*{start nodes}
    $v = rand(\altmathcal{R})$;
    let $\altmathcal{T} = N^+(v)$;
    $p$ = [$v$]\;
    tried = \{\};
    visited = \{\}

    \While{$v \not= v_{dt}$ \textbf{and} $|\altmathcal{T}| >$ 0 \textbf{and} $|tried|  < |\altmathcal{T}|$}{
        $\altmathcal{T}$ = $\altmathcal{T}$ / tried\; 
        \uIf{attack$\_$strategy == $RWE$}{
            $v \leftarrow X_1$\; 
        }
        \uElseIf{attack$\_$strategy == $RE$}{
            $\bm{r}$ = $\bm{r}_o(\altmathcal{T})$;
            $v \leftarrow X_2$\; 
        }
        \uElseIf{attack$\_$strategy == $DE$}{
            $\bm{r}$ = $\bm{\delta}_o(\altmathcal{T})$;
            $v \leftarrow X_3$\;
        }
        $\altmathcal{T}$ = $\altmathcal{T}$ $\cup$ tried\; 
        
        \uIf{$Valid(v)$ \textbf{and} $v\not\in$ visited}{ 
            tried = \{\}\;
            $\altmathcal{T} = \altmathcal{T} \setminus \{v\} \cup N^{+}(v)$\;
            $\bm{p}$ += $v$;
            visited += $v$\; 
        }
        \Else{
            tried $\cup$ $v$\;
        }
        
    }
    
    Return $\bm{p}$\;
    
    \caption{Lateral Attack Modeling}
    \label{algorithm:attack}
\end{algorithm2e}

%% file: Algorithms/AnomalyShield.tex
\begin{algorithm2e}[h]
\footnotesize
    \DontPrintSemicolon
    \KwIn{Adjacency matrix $\bm{A}$, anomaly vector $\bm{a}$, and vaccination budget $k$}
    \KwResult{a set $\altmathcal{S}_k$ with $k$ nodes}
    
    Compute first eigenvalue $\lambda$ and corresponding eigenvector $\bm{u}$ of $\bm{A}$\;
    
    $\bm{c}$ = $\bm{A}$ * (a * u)\;
    score = $\bm{c}$ * $\bm{u}$

    \For{iter = 1 to $k$}{
        $v$ = argmax$_{i}$ score($i$), \text{add $v$ to set $\altmathcal{S}$}\;
        score($v$) = -1\;
     }
     
     return $\altmathcal{S}$\;

    \caption{\method{}}
    \label{algorithm:anomaly_shield}
\end{algorithm2e}

%% file: Sections/04-Experiments.tex
\section{Experiments}

\subsection{Experimental Setup}
\label{sec:experimental-setup}
All experiments are conducted on three real authentication graphs, collected over 30 days (statistics in Table~\ref{table:statistics}). 
Two graphs are from Microsoft: anonymized enterprise networks $G_s$ and $G_l$;
and one is from Los Alamos National Lab \cite{akent-2015-enterprise-data}: open-sourced network $G_{lanl}$. 
For each attack strategy and hygiene level, 
we strive to collect 200 unique attack paths for 50 credential distributions $\bm{d}\in D$.
These parameters are determined based on the available 2-week computation budget for data collection. 
Certain combinations of $G$ and $\bm{d}$ have a high rate of attack failure; 
we terminate the collection process at 10,000 failed attempts, collecting as many as possible.

\input{Tables/Graph_Statistics.tex}

\subsection{Network Vulnerability Analysis}\label{results:vulnerability}
In Table~\ref{table:attack_paths}, we summarize the  first  experimental  results on  network  vulnerability  to  lateral  attack 
by analyzing the attack strategies
\textit{Rank-Explore} (RE), 
\textit{Degree-Explore} (DE), and 
\textit{RandomWalk-Explore} (RWE)
(discussed in Sect.~\ref{section:simulation}).
For each strategy, we average the attack path length across all credential distributions.
We compute the network vulnerability statistics using Eq.~\ref{eq:vulnerability_1}---hygiene-specific $L(G,h)$; and Eq.~\ref{eq:vulnerability_2}---whole-network $L(G)$ from Section~\ref{section:vulnerability}.    
We identify multiple key insights:

\begin{enumerate}[topsep=1mm, itemsep=0mm, parsep=1mm, leftmargin=*]

\item \textbf{Informed Strategies Lead to Quicker Attacks} 
The 
RE
and 
DE
strategies produce shorter paths in general, 
compared to 
RWE. 
This is expected, as prior knowledge should help the attacker reach the domain controller in less time. Also, adversaries likely prefer shorter attack paths, which leaves smaller footprints 
for anomaly systems to detect. 

\item  \textbf{Improving Hygiene Reduces Vulnerability} Increasing network hygiene ($h_1\rightarrow{}h_2\rightarrow{}h_3$) causes longer attack paths (or none at all) and generally reduces vulnerability (e.g., for $G_s$ and $G_l$).
On graph $G_s$, the highest hygiene level $h_3$ critically reduces high-level admin credentials, 
significantly improving network robustness (vulnerability reduced to $0$).
Such findings can empower IT admins to develop robust user access credential policies.

\item \textbf{Linking Topology to Network Vulnerability}
Networks that are well-connected are more vulnerable to lateral attack (e.g., $G_{lanl}$, with higher average clustering coefficient and node degree).
This is expected, due to increased lateral movement opportunities. 
Relatedly, improving network hygiene level in such a well-connected network does not seem to reduce network vulnerability. 

\end{enumerate}

\input{Tables/Attack_Paths.tex}

\subsection{Defense Strategy Analysis}\label{results:vaccination}
\begin{figure}[!t]
\centering
\includegraphics[width=\linewidth]{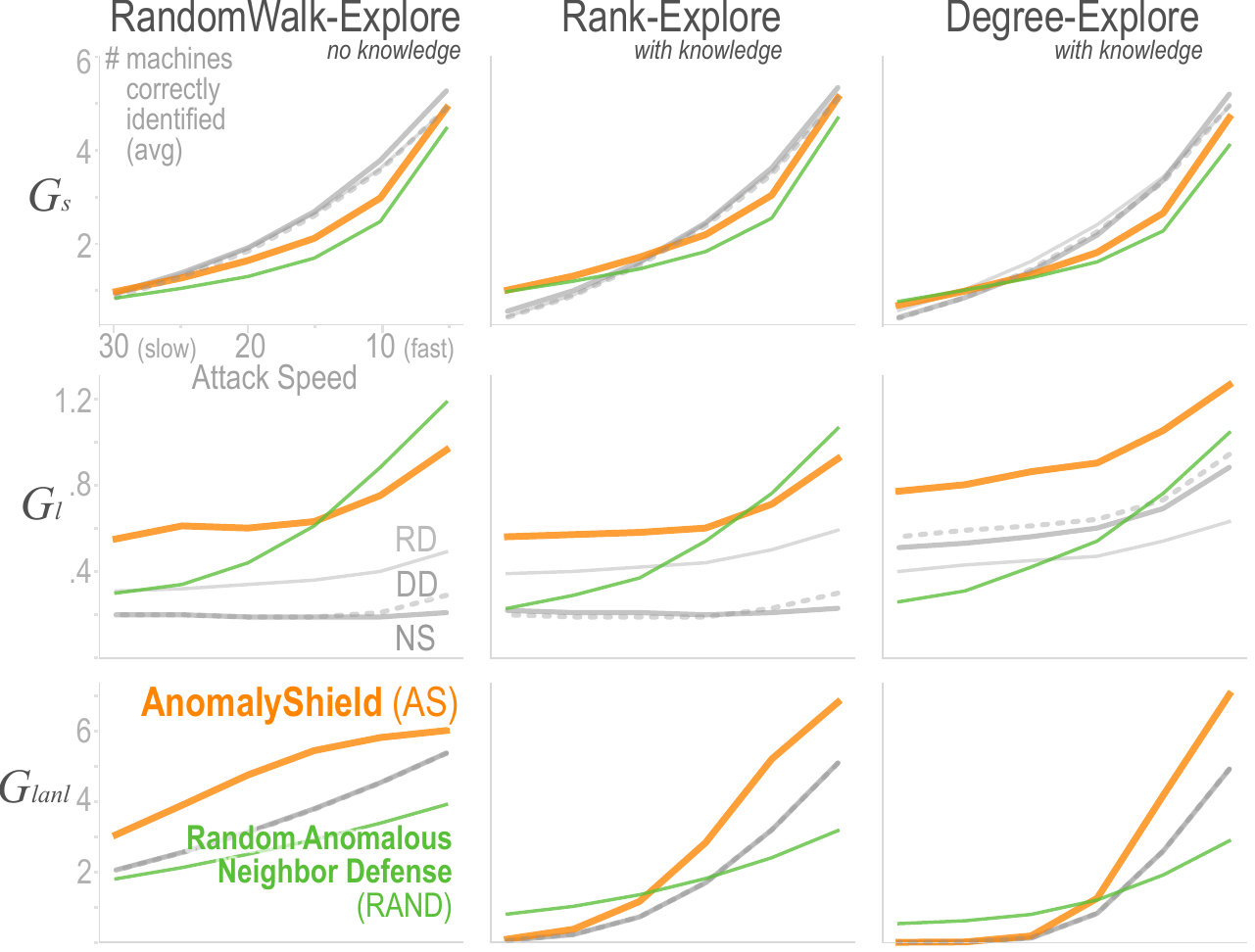}
\caption{
Each defense strategy is compared on three graphs and attack strategies, where \method{} performs well across a majority of application scenarios.}
\label{fig:vaccination_results}
\end{figure}

We report the first results for identifying machines at-risk to lateral attack, evaluating each defense strategy proposed in Section~\ref{sec:defense}. 
We measure the success of each strategy by its ability to predict attacker movement.
That is, given graph topology $G$ and suspected lateral attack movement $\bm{p}^i,...,\bm{p}^{0}$, 
predict attack activity at $\bm{p}^{i+1}$ (each $\bm{p}^i$ is a sequence/path of suspected machines traversed by the attacker). 
Formally, we intersect the predicted \textit{at-risk machines} $\altmathcal{S}_k$ with $\bm{p}^{i+1}$.
Since the defender likely monitors the domain controller, we exclude it from $\altmathcal{S}_k$. 
We repeat this process for each sub-path (except $\bm{p}^0$) and average over all attack paths. 
Figure~\ref{fig:vaccination_results} shows every combination of attack and defense strategy, with budget $k$=8 and hygiene $h_2$, which provide representative results. We identify multiple key insights:

\begin{enumerate}[topsep=1mm, itemsep=0mm, parsep=1mm, leftmargin=*]
    \item \textbf{\method{} as Effective General Defense}
    \textcolor{as-color}{\method{}}
    generally performs well (identified more machines) across:
    network topology (rows in figure),
    adversary's prior knowledge (columns),
    and attack speed (horizontal axes).
    We believe this is because \method{} focuses on high-centrality machines with anomalous neighbors, combining desirable attributes from static and dynamic methods.

    \item \textbf{Similar Effectiveness in Small Graphs} 
    All strategies perform similarly in small graph $G_s$ (first row),
    since fewer machines exist for monitoring.
    
    \item \textbf{Large Graphs Require Informed Defense} 
    Uninformed defense strategy \textcolor{rand-color}{\textbf{RAND}} 
    is significantly less effective in the large graph $G_{lanl}$ (last row), especially when encountering faster attacks.
    This could be explained by the need for intelligent decision making in the presence of many options.
\end{enumerate}

%% file: Tables/Graph_Statistics.tex
\begin{table}[t]
\setlength{\tabcolsep}{5pt}
\centering
\footnotesize
\centering
 \begin{tabular}{llrrrrr} 
 \toprule
 
  Graph & Source & $|V|$ & $|E|$& $\rho$ & $C$ & $\delta_{avg}$ \\
 \midrule
 \textbf{$G_s$} & Microsoft & 100 & 279 & 0.028 & 0.23 & 5.58 \\ 
 \textbf{$G_l$} & Microsoft & 2,039 & 3,853 & 0.001 & 0.26 & 3.78 \\ 
 \textbf{$G_{lanl}$} & LANL & 14,813 & 223,399 & 0.001 & 0.62 & 30.16 \\

\bottomrule
\end{tabular}

\caption{Graph Statistics. $\rho$: graph density, $C$: average clustering coefficient, $\delta_{avg}$: mean node out-degree.}
\label{table:statistics}
\end{table}

%% file: Tables/Attack_Paths.tex
\begin{table}[t]
\footnotesize
\setlength{\tabcolsep}{5pt}
\centering

 \begin{tabular}{ll|rrr|rr} 
 \toprule
 
& & \multicolumn{3}{c}{{\textbf{Avg. Path length}}} & \multicolumn{2}{c}{{\textbf{Vulnerability}}} \\
\cmidrule(l){3-5} \cmidrule(l){6-7} 
    
  \textbf{Graph} & \textbf{Hygiene} & \textbf{RE} & \textbf{DE} & \textbf{RAND} &  \textbf{$L(G, h)$} & \textbf{$L(G)$} \\

 \midrule
 \multirow{3}{*}{\textbf{$G_s$}}
 & $h_1$ & 19 & 19 & 25 &  .773 &  \\ 
 & $h_2$ & 49 & 39 & 39 & .801 & .525 \\ 
 & $h_3$ & 0 & 0 & 0 & 0 &  \\ 
 \addlinespace[0.2cm]
 
 \multirow{3}{*}{\textbf{$G_l$}}
 & $h_1$ & 33 & 36 & 46 & .005 &  \\ 
 & $h_2$ & 63 & 63 & 68 & .006 &  .005 \\ 
 & $h_3$ & 133 & 139 & 139 & .004 &  \\ 
 \addlinespace[0.2cm]
 
 \multirow{3}{*}{\textbf{$G_{lanl}$}}
 & $h_1$ & 22 & 18 & 45 & .967 &  \\ 
 & $h_2$ & 88 & 128 & 90 & .981 & .976\\ 
 & $h_3$ & - & - & 249 & .981 & \\ 
 
\bottomrule
\end{tabular}

\caption{Vulnerability Statistics. 
Statistics excluded for $G_{lanl}$ strategies RE and DE in $h_3$ as computation exceeded budget (Sect.~\ref{sec:experimental-setup}).
}
\label{table:attack_paths}
\end{table}

%% file: Sections/07-Conclusion.tex
\section{Conclusion}
We present  \model{}, the first framework that systematically quantifies network vulnerability to lateral attacks and identifies at-risk devices.
\model{} models lateral attacks on enterprise networks using attack strategies developed with Microsoft.
We formulate network vulnerability as a novel Monte-Carlo method and
propose a suite of five fast graph mining techniques, including the novel \method{} method, to identify at-risk machines.
Using real data, we demonstrate \model{}'s unique potential to empower IT admins to develop robust user access credential policies.


%% file: Sections/08-Acknowledgments.tex
\section{Acknowledgements}
This work was in part supported by the NSF grant IIS-1563816 and GRFP (DGE-1650044).